\documentclass[10pt,leqno]{amsart}
\renewcommand{\vec}[1]{\boldsymbol{#1}}         
\newcommand{\pd}{\partial}   
\newcommand{\pfrac}[2]{\frac{\pd #1}{\pd #2}}

\usepackage[numbers]{natbib}
\usepackage{float}
\usepackage{tikz}
\usetikzlibrary{external}
\usepackage[utf8]{inputenc}
\usepackage{subfig}
\usepackage{pgfplots}
\pgfplotsset{compat=newest}
\usepgfplotslibrary{groupplots}
\usepgfplotslibrary{dateplot}
\newlength\ah
\newlength\aw
\newlength\ahdp
\newlength\awdp
\usepackage{caption}
\usetikzlibrary{patterns}

\begin{document}

\let\WriteBookmarks\relax
\def\floatpagepagefraction{1}
\def\textpagefraction{.001}

\title[Localized kernel gradient correction for SPH simulations of water waves]{Localized kernel gradient correction for SPH simulations of water wave propagation}

\author[L.J. Schulze, V. Zago, G. Bilotta, R.A. Dalrymple]{Lennart J. Schulze$^{1,*}$\and Vito Zago$^1$\and Giuseppe Bilotta$^2$\and Robert A. Dalrymple$^1$}
\begingroup
\renewcommand\thefootnote{}\footnotetext{
  \\$^{1}$: Dept. of Civil and Environmental Engineering, Northwestern University, 2145 Sheridan Road, Evanston, IL 60208, USA\\
  $^{2}$: Osservatorio Etneo, Istituto Nazionale di Geofisica e Vulcanologia, Catania, Italy\\
  $^*$: Corresponding author, \texttt{lschulze@mpi-cbg}\\Note: Parts of this manuscript have been previously presented in the internal proceedings of the SPHERIC 2022 International Workshop, June 7-9 2022, Catania, Italy.
}
\addtocounter{footnote}{0}
\endgroup

\begin{abstract}
Basic Smoothed Particle Hydrodynamics (SPH) models exhibit excessive, numerical dissipation in the simulation of water wave propagation. This can be remedied using higher-order approaches such as kernel gradient correction, which introduce additional computational effort. 

The present work demonstrates, that the higher-order scheme is only required in a limited part of the water wave in order to obtain satisfying results. The criterion for distinguishing particles in need of special treatment from those that do not is motivated by water wave mechanics. Especially for deep water waves, the approach potentially spares large amounts of computational effort. The present paper also proposes a remedy for issues of the kernel gradient correction occurring at the free surface. 

Satisfying results for the proposed approach are shown for a standing wave in a basin and a progressive wave train in a long wave tank.
\end{abstract}
\maketitle
\section{Introduction}
Water wave propagation constitutes a problem of great importance for a wide range of scientific topics in coastal engineering such as wave shoaling or the propagation of tsunamis. Since the applicability of theoretical solutions is limited and experimental research can be time-consuming and costly, numerical approximations are gaining importance. Smoothed Particle Hydrodynamics (SPH) is a meshless Lagrangian method which is particularly well-suited for the simulation of water wave propagation due to its straightforward interface tracking. However, it suffers from excessive numerical dissipation during the simulation of low-viscosity media. Without efficient and stable remedies, this circumstance limits the potential applicability of SPH.

Simulating gravity waves with SPH has been the subject of several publications \cite{monaghan_1994, dalrymple_2006, trimulyono_2019, Chang_2017, guilcher_2007, gao_2012, wen_2018, omidvar_2015, zago_2021}. The literature discusses wave propagation with a range of Reynolds numbers of waves. However, low-viscosity media pose a particularly challenging problem due to the small amount of energy dissipation \cite{guilcher_2007}. For this reason, waves associated with high Reynolds numbers such as water waves remain underrepresented. 

The comprehensive study by Colagrossi et al. \cite{Colagrossi_2013} concludes that basic SPH models are capable of accurately predicting the attenuation process in viscous standing waves. The number of neighbors per particle is the decisive factor, which assures convergence towards analytical results with growing spatial resolution for a spectrum of Reynolds numbers. Colagrossi et al. link the excessive attenuation to the generation of spurious vorticity. While the choice of the kernel function can improve this at least to some extent \cite{Colagrossi_2013, Macia_2011, Chang_2017, zhang_2018, omidvar_2015}, the simulation of waves with large Reynolds numbers quickly demands an unfeasible amount of computational effort due to the required number of neighbors and spatial resolution.

Numerous studies have developed a range of schemes in order to relax the criteria discussed by Colagrossi. Chang et al. \cite{Chang_2017} employ a sixth-order kernel with a large smoothing length. This approach is simple and provides accurate results. However, although the sixth-order kernel appears to be a well-suited choice, the required number of neighbors per particle still calls for great computational effort, which increases significantly in 3D. Riemann solvers \cite{guilcher_2007, gao_2012, omidvar_2015} have also proven to result in an improvement in wave propagation with SPH, but those often involve a complex implementation and do not account for any viscous terms without further, special treatment.

In addition to Riemann solvers, Guilcher \cite{guilcher_2007} and Gao \cite{gao_2012} incorporate forms of kernel gradient correction. Wen et al. \cite{wen_2018} combine the latter with an additional diffusive term in the continuity equation. Kernel correction schemes are based on the original propositions by Randles and Libersky \cite{randles_libersky_1996} and Johnson and Beissel \cite{johnson_beissel_1996}, which also appear in the Corrective Smoothed Particle Method (CSPM) by Chen and Beraun \cite{chen_1999}. Corrective methods are known to enhance the consistency of the kernel gradient \cite{liu_2010} and have shown to significantly improve results in the simulation of water wave propagation without any extensive increase of the smoothing length.

Obtaining the coefficients necessary for the multiplicative correction of the kernel gradients as in \cite{randles_libersky_1996} involves a summation over the neighborhood of a particle as well as the inversion of a matrix, leading to additional computational efforts. Furthermore, the application of the corrective scheme is known to cause errors in some geometrical configurations \cite{xiao_2020}. These errors potentially impede the stability of the underlying framework, as the matrix can become singular in certain particle distributions. A further drawback of kernel gradient correction are the asymmetric interactions in the momentum equation, as the gradients are corrected with particle-specific coefficients \cite{oger2007}. This leads to a loss of conservation of momentum, which plays a crucial role in wave propagation.

Zago et al. \cite{zago_2021, zago_spheric_2021} solved the issues of momentum conservation by proposing a symmetric kernel gradient correction scheme. The method also addresses the stability issues associated with kernel gradient correction by introducing a threshold dependent on the local particle distribution. Whilst the obtained results are highly accurate and satisfying, the method depends on a careful choice of the introduced threshold, and still involves a computational overhead, which leads to increased required simulation times.

In order to lessen the computational overhead, this work shall consider water wave mechanics: The dynamics of deep water waves exhibit an exponential decay over the water depth. If a depth of half a wave length is surpassed, virtually no water particle movement occurs anymore. The impact of an application of higher-order SPH behaves in a similar way, yielding great improvements for areas close to the surface, and negligible differences when applied in great depths. Depending on the water wave considered, the gain in efficiency can be great. By excluding the part of the domain that lies in great depths from the application of higher-order SPH, the computational demand is reduced, without significant degradation of the results.

Further, this work proposes an alternative remedy to the stability issues, which frees the framework from a heuristic threshold. This is done by weighting the gradients according to a quality measure of the local support of a particle, namely the kernel support.

The present work addresses mentioned propositions in the context of both a standing wave and a progressive wave train.

\section{Linear wave theory}\label{sec:linwavetheo}
In order to motivate the method proposed within the present work, insights from the linear wave theory are briefly summarized. Linear wave theory derives solutions for the propagation of gravity waves from the mass continuity equation for a velocity potential under the assumption of inviscid, incompressible and irrotational flow. The applicability of the theory is confined to the propagation of linear waves, which excludes wave breaking or other nonlinear phenomena. 

\begin{figure}[h!]
    \centering
    \includegraphics[]{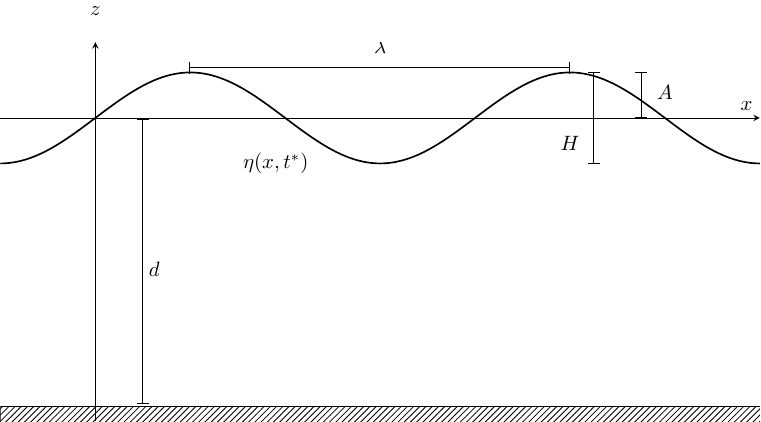}
    \caption{A propagating wave.}
    \label{fig:wave}
\end{figure}

Linear waves are classified according to their ratio of water depth $d$ to wave length $\lambda$ as shallow water for $\tfrac{d}{\lambda}<0.05$ and deep water for $\tfrac{d}{\lambda}>0.5$. The realm between shallow and deep water is referred to as intermediate depth. Deep water waves with small height $H$ compared to the wave length $\lambda$, and shallow water waves with small $H$ compared to the water depth $d$ qualify as linear waves. For these waves, linear theory provides good theoretical approximations.
Figure \ref{fig:wave} displays the surface elevation $\eta$ for a time $t^*$ and further important parameters of a wave. The points of maximum and minimum value of $\eta\left(x,t\right)$ are called crests and troughs, respectively.
For the sake of brevity, the derivation of the solutions of linear theory is omitted. The interested reader is referred to Dean and Dalrymple \cite{dean1991water}. The following equations refer to a coordinate system as depicted in Figure \ref{fig:wave}, the origin lies within the still water level of the wave and the $x$-axis points towards the direction of propagation, and all properties remain constant throughout the $y$-direction.

For a known wave height $H$ and wave length $\lambda$, the velocities of fluid particles within a progressive wave are obtained as
\begin{align}
u_x=\pfrac{\phi}{x}=\frac{H\text{g}k}{2\omega}\frac{\cosh{\left(k\left(d+z\right)\right)}}{\cosh{\left(kd\right)}}\cos{\left(kx-\omega t\right)},\label{u_x}\\
u_z=\pfrac{\phi}{z}=\frac{H\text{g}k}{2\omega}\frac{\sinh{\left(k\left(d+z\right)\right)}}{\cosh{\left(kd\right)}}\sin{\left(kx-\omega t\right)},\label{u_z}
\end{align}
in which the wave number is denoted by $k=\frac{2\pi}{\lambda}$. The angular frequency of the wave $\omega$ can be determined by the dispersion relationship,
\begin{equation}\label{eq:dispersionrelation}
\omega^2=\text{g}k\tanh{\left(kd\right)},
\end{equation}
in which the gravitational acceleration is $\text{g}\approx9.81$. Further, the period of the wave can be obtained by considering $T=\frac{2\pi}{\omega}$ and the phase velocity of the wave results from the wave length and frequency as $u_p=\frac{\omega}{k}$.

The hyperbolic terms in the velocities \eqref{u_x} and \eqref{u_z} can be expressed with exponential functions as $\sinh{x}=\tfrac{\text{e}^x-\text{e}^{-x}}{2}$ and  $\cosh{x}=\tfrac{\text{e}^x+\text{e}^{-x}}{2}$, such that they result in
\begin{align}
    \frac{\cosh{\left(k\left(d+z\right)\right)}}{\cosh{\left(kd\right)}}=\text{e}^{kz}\frac{1+\text{e}^{-2k(d+z)}}{1+\text{e}^{-2kd}},\\
    \frac{\sinh{\left(k\left(d+z\right)\right)}}{\cosh{\left(kd\right)}}=\text{e}^{kz}\frac{1-\text{e}^{-2k(d+z)}}{1+\text{e}^{-2kd}}.
\end{align}
In case those terms are considered for velocities close to the surface for deep water waves, i.e. $d\xrightarrow{}\infty$, they both yield $\text{e}^{kz}$. As the other terms in \eqref{u_x} and \eqref{u_z} are either constant or bounded, the velocities for fluid particles in deep water waves display an exponential decay with respect to the distance from the still water level. At a depth of $z=-\tfrac{\lambda}{2}$, the magnitude of the particle velocities are only about four percent of those at the surface. As the kinetic energy is proportional to the square of the velocities, it is localized almost entirely within the distance of $z=-\tfrac{\lambda}{2}$ from the still water level of linear deep water waves. The same observation can be made for deep water standing waves.

Linear wave theory provides further insights regarding the pressure field within water waves. One contribution to the pressure field, the well-known hydrostatic pressure, is also present in still waters. The second contribution is dependent on the water surface elevation $\eta$ and is referred to as the dynamic pressure. The pressure field yields
\begin{equation}\label{eq:pressureconti}
    p(x,z,t)=-\rho\text{g}z+\rho\text{g}\eta\left(x,t\right)\frac{\cosh\left(k\left(d+z\right)\right)}{\cosh\left(kd\right)}.
\end{equation}
For deep water waves, the hyperbolic terms in the dynamic pressure term causes the same decay with particle depth as displayed in the velocity field.

\section{An efficient and accurate numerical approach}
The present work simulates water wave propagation based on a discretized version of the weakly incompressible Navier-Stokes equations using SPH, which shall briefly be introduced in the following.

\subsection{Smoothed Particle Hydrodynamics}
The SPH approximation is a meshless Lagrangian scheme which represents a fluid as a finite set of discrete collocation points, referred to as particles. These particles move with the fluid and advect certain properties of it, such as the density $\rho$, the mass $m$, or the pressure $p$. Advantages of SPH include the exact simulation of pure advection, its adaptivity and stability properties. 

The method represents an arbitrary field $f$ as its convolution with a kernel function $W$, which has similar properties as the Dirac delta distribution. The integral is then generally approximated using a quadrature with $N_p$ discrete quadrature points, the particle positions. The accuracy of the method is mainly determined by the smoothing error and the discretization error. The former is related to the smoothing length $h$ of the kernel function, and the latter is related to the particle spacing $\Delta p$. 

Within the present paper, the kernel function $W$ is the quintic Wendland kernel \cite{wendland_1995}, as it has shown to be well-suited for the application of free-surface simulations \cite{Macia_2011}. For derivative fields, it can be beneficial to avoid potential singularities when particles approach each other by formulating the derivative of the kernel function as
\begin{equation}
F(r, h) = \frac{1}{r} \frac{\partial W(r, h)}{\partial r}.
\end{equation}
The simulation of low-viscosity media such as water using a standard SPH approach is characterized by a great amount of excessive dissipation. As the conservative corrected SPH (CCSPH) has proven to avoid numerical dissipation in water wave propagation \cite{zago_2021}, the method of the present paper will employ a similar variant of the method for the discretization of the Navier-Stokes equations.

Analogous to \cite{zago_2021}, the approximation of the continuity equation introduces an artificial density diffusion term according to \cite{molteni_colagrossi_2009} and yields
\begin{equation}\label{eq:part_continuity}
\frac{D\rho_i}{Dt} = \sum_{j} \vec u_{ij}\cdot \vec r_{ij} F_{ij} m_j + \xi h c_0 \sum_j \Psi_{ij} F_{ij} m_j,
\end{equation}
in which the velocity difference is denoted by $\vec u_{ij} = \vec u_i-\vec u_j$, the versor from particle $i$ to $j$ is denoted by $\vec r_{ij}$, and $c_0$ is the speed of sound of the fluid. The density diffusion term is given by
\begin{equation} \label{eq:molteniColagrossi}
\Psi_{ij} = \left\{ \begin{array}{ccl}
	2\left(\frac{\rho_j}{\rho_i} - 1 \right) & \text{if} & \frac{|p_i - p_j|}{\rho_i g |z_i - z_j|}>1\\
	0 & \text{otherwise.}
\end{array} \right.
\end{equation}
The regularization coefficient is set to $\xi=0.1$ within this work.

The momentum equation is discretized as 
\begin{equation}\label{eq:corr_part_NS}
\frac{D\vec u_i}{Dt} = \sum_{j}
\left( \frac{p_i}{\rho_i^2} +
\frac{p_j} {\rho_j^2} + \Pi_{ij}\right )
 \mathbf{B}_{ij} \vec x_{ij} F_{ij} m_j+ \vec g,
\end{equation}
in which the gravitational force is modeled in $\vec g$. The artificial viscosity term according to \cite{monaghan_2005} introduces reads
\begin{equation}\label{eq:artvisc}
\Pi_{ij} =  \left\{ \begin{array}{ccl}
- \frac{\alpha \, h \, c_0}{\rho_j} \left(\frac{\vec u_{ij} \cdot \vec x_{ij}}{|\vec x_{ij}|^2 + \epsilon h^2}\right) & \text{if} & (\vec u_{ij} \cdot \vec x_{ij}) >1\\
0 & \text{otherwise.}
\end{array} \right.
\end{equation}
The constant preventing singularities is chosen as $\epsilon=0.01$, and the artificial viscosity coefficient $\alpha$ can be related to an equivalent kinematic viscosity \cite{monaghan_2005} as
\begin{equation}\label{eq:nuofalpha}
\nu = \frac{hc_0\alpha}{10}.
\end{equation}
The system is complemented with the equation of state \cite{cole_48}, by which the pressure of a particle $p_i$ is linked to its density $\rho_i$:
\begin{equation}\label{eq:eos}
p_i(\rho_i)=c_0^2 \,\frac{\rho_0}{\gamma} \left[\left(\frac{\rho_i}{\rho_0}\right)^\gamma-1\right].
\end{equation}

Finally, according to \cite{zago_2021} the symmetric corrective coefficients in \eqref{eq:corr_part_NS} read:
\begin{equation}
    \vec B_{ij} = \left(\frac{1}{2}(\vec A_i + \vec A_j)\right)^{-1},
\end{equation}
with $\vec A_i$ computed as
\begin{equation}\label{eq:matrix_A}
\mathbf{A}_i= \sum_j \nabla W_{ij} \otimes (\vec{x}_j - \vec{x}_i) V_j = \sum_j
\begin{pmatrix}
 x_{ij}\frac{\partial W_{ij}}{\partial x}& y_{ij}\frac{\partial W_{ij}}{\partial x}& z_{ij}\frac{\partial W_{ij}}{\partial x}\\
 x_{ij}\frac{\partial W_{ij}}{\partial y}& y_{ij}\frac{\partial W_{ij}}{\partial y}& z_{ij}\frac{\partial W_{ij}}{\partial y}\\
 x_{ij}\frac{\partial W_{ij}}{\partial z}& y_{ij}\frac{\partial W_{ij}}{\partial z}& z_{ij}\frac{\partial W_{ij}}{\partial z}
\end{pmatrix}
V_j,
\end{equation}
in which $V_j=\tfrac{m_j}{\rho_j}$ is the volume associated with the particle $j$. In the following, two alterations to the presented scheme are proposed, which concern the classically associated issues with kernel gradient correction.
\subsection{Stability issues and weighting factors}\label{sec:weightingfactors}
Kernel gradient correction is known to cause instabilities in some geometrical configurations \cite{xiao_2020}. In those cases, the corrective matrix may become ill-conditioned. A basic particle distribution, which causes $\mathbf{A}_i$ to become singular, is easily imagined as a single line of perfectly aligned particles. For this configuration, the differences in  coordinates vanish for all particles within the local neighborhoods. This in turn, causes matrix entries of $\mathbf{A}_i$ to vanish, resulting in a zero determinant.

An exact alignment of particles without any deviations is unlikely. But, splashes caused by breaking waves or a very gentle slope of a beach that is modeled by geometrical planes, without any boundary particles that contribute to the support of fluid particles, can lead to similar distributions and ill-conditioned corrective matrices. The inversion and application of those matrices lead to spurious and often very large forces in the momentum equation, resulting in an overall less stable framework. True singularities require thresholds on the determinant of the corrective matrix \eqref{eq:matrix_A} as done in \cite{zago_2021}, or the distance from boundaries. However, other irregular particle distributions can cause instabilities and spurious effects as well.

As an alternative remedy free of any parameters, the corrected kernel gradients shall be weighted by a measure of the quality of the local particle distribution. For this purpose, the support of the particle $i$ is considered, which shall be denoted by $w_i$:
\begin{equation}\label{eq:sphsupport}
    w_i=\sum_{j=1}^{n_p}V_jW_{ij}.
\end{equation}
If $w_i\approx1$, the normalization property of the kernel function holds not only in a continuous, but also in a discrete sense and particle $i$ has a regular support. If $w_i$ becomes larger, it is located in a densely packed region. If the support becomes smaller, it lacks surrounding particles and may be close to the surface of the fluid, for example.

As the averaged corrective coefficients are associated with particle pairs $ij$, the weighting factor should also take both neighborhoods into account. This is done by computing the arithmetic mean
\begin{equation}\label{eq:sphwmean}
    w_{ij}=\frac{w_i+w_j}{2}.
\end{equation}
The kernel gradients of particle pairs lacking support are effectively scaled down if weighted by $w_{ij}$. Especially in the proximity of the free surface, this will take place and limit the influence of spurious corrective coefficients. Note, that due to the self contribution $j=i$ in Eq. \eqref{eq:sphsupport} and the kernel function always taking on positive values, the support of a particle never vanishes and gradients cannot be erased due to the weighting.

Introducing the weighting factor to the kernel gradient correction yields a stabilized symmetric kernel gradient correction, namely
\begin{equation}
    \vec B_{ij} = w_{ij}\left(\frac{1}{2}(\vec A_i + \vec A_j)\right)^{-1} = (w_i+w_j)(\vec A_i + \vec A_j)^{-1}.
\end{equation}
\subsection{Computational effort and correction of a restricted subset}\label{sec:correctedsubsets}
The computation and application of the corrective matrix involves additional computational effort compared to basic SPH frameworks if the smoothing length is kept constant. This effort is mainly impacted by the assembly of $\mathbf{A}_i$, which takes place in the local neighborhood of each particle.

To lessen this computational burden, the particles within the domain are separated into a corrected subset and an uncorrected subset depending on the contribution of a particle to the overall dynamics of the water wave. The latter subset will employ basic SPH gradient approximations, and hence spare the computational expense of the computation of corrective coefficients. This principle of employing different gradient formulations for different particles, is similar to the approach of Jiang \cite{jiang_2011}, in which the higher-order corrective scheme Modified Smoothed Particle Hydrodynamics (MSPH) \cite{zhangbatra2004} is applied only to so called exterior particles.

In the following, the set containing all particles $P_i$ for $i\in\left\{1,...,n_p\right\}$ in the domain shall be denoted by $\mathcal{P}$, whereas the set containing particles for which the corrective coefficients \eqref{eq:matrix_A} are computed shall be \begin{equation}
    \mathcal{C}\subseteq\mathcal{P}.
\end{equation}
A further subset $\mathcal{I}$ is defined as the complement
\begin{equation}
    \mathcal{I}=\mathcal{P}\setminus\mathcal{C},
\end{equation}
which contains all particles at which the corrective matrix is not computed but simply set to identity. 

In order to determine a sensible rule to distinguish the members of $\mathcal{C}$ from the members of $\mathcal{I}$, the approximations from linear wave theory shall be considered. As highlighted before, the velocity field \eqref{u_x}-\eqref{u_z} exhibits an exponential decay in magnitude with respect to the distance from the still water level. In case a deep water wave is discretized using SPH, this circumstance results in potentially large amounts of particles without any significant contribution to the dynamics of a water wave. Vice versa, particles closer to the surface are essential to the evolution of waves and their kinetic energy, and thus may require special treatment. 

\begin{figure*}[h!]
\centering
\includegraphics[]{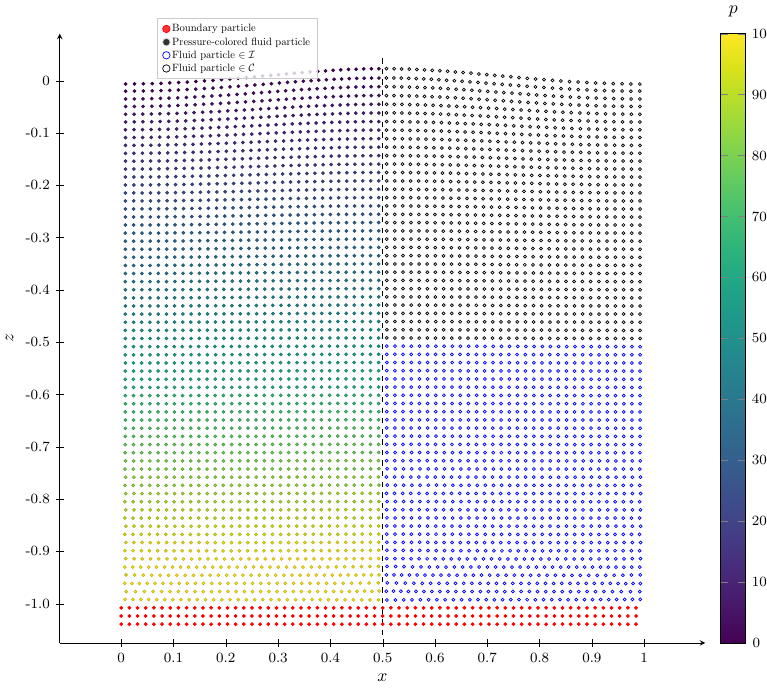}
\put(0,320){$\times 10^{4}$}
\caption{Standing wave particle distribution computed by regularized CCSPH and $\chi=1/2$ at time $t=0.2$. For the sake of illustration, a large interparticle spacing of $\Delta p=1/64$ has been chosen. The left half of the domain displays the pressure field of the fluid particles, whereas the right half of the domain exhibits the different subsets $\mathcal{I}$ and $\mathcal{C}$ that result from the pressure criterion.}
\label{fig:standingwavechihalf}
\end{figure*}

These considerations shall be incorporated in the criteria for the definition of the aforementioned subsets. Note that in the following, the $z$-coordinate refers to the system depicted in Fig. \ref{fig:wave}. The sets yield
\begin{align}
    \mathcal{C}=\left\{P_i|z_i\geq-\chi\lambda\right\},\\
    \mathcal{I}=\left\{P_i|z_i<-\chi\lambda\right\},
\end{align}
in which the parameter $\chi$ determines the range of depth of particles for which correction coefficients are computed. A possible value is $\chi=1/2$, as this would cover an area which contributes more than 95 percent of kinetic energy according to linear wave theory. Figure \ref{fig:standingwavechihalf} displays such subsets for the example of a standing wave.

Since surface detection routines are not always present in frameworks and potentially hinder optimal performance, an alternative distinguishing criterion to quantify the depth of a particle is proposed and employed within the context of this thesis. Linear wave theory approximates the pressure field in water waves through the static and dynamic pressure terms in Eq. \eqref{eq:pressureconti}. For the purpose of deriving an alternative criterion, the dynamic term is omitted and the pressure of a particle is compared to the hydrostatic pressure field:
\begin{equation}\label{eq:sphzassumption}
    z_i\approx-\frac{p_i}{\rho_0\text{g}}.
\end{equation}
Inserting the condition $z_i\geq-\chi\lambda$ and solving for $p_i$ yields a pressure condition for members of $\mathcal{C}$, namely
\begin{equation}\label{eq:sphpcondlambda}
    p_i\leq\chi\lambda\rho_0\text{g}.
\end{equation}
Eq.\eqref{eq:sphpcondlambda} demands members of $\mathcal{C}$ to have a smaller pressure than the hydrostatic pressure at the initially determined depth. The pressure $p_i$ can further be substituted using an inverted form of the equation of state \eqref{eq:eos}, such that a condition on the density of a particle $\rho_i$ is obtained.

For small-amplitude waves, the impact of the dynamic term is negligible compared to the hydrostatic pressure evaluated at $-\chi\lambda$.

The proposal to distinguish members of $\mathcal{C}$ from members of $\mathcal{I}$ based on the SPH pressure field demands a certain degree of smoothness of the pressure field. For this reason, the density diffusion term in Eq. \eqref{eq:molteniColagrossi} plays an important role.

Naturally, the amount of spared computational effort strongly depends on the wave parameters and the domain geometry. Within deep water, $d>\tfrac{\lambda}{2}$, there is a potentially large performance gain. Employing the proposed division in subsets for shallow water waves, however, resorts to all particles becoming members of $\mathcal{C}$ and an unaltered performance. The choice of $\chi$ is influenced by the surrounding framework. If, for example, a larger smoothing factor $\tfrac{h}{\Delta p}$ is chosen, $\chi$ can take on smaller values.

For the sake of proposing a generic solution, particles that have a boundary in their support, or are part of the boundary themselves, are assigned to subset $\mathcal{I}$.

After discretizing the Navier-Stokes equations in terms of space, an ordinary differential equation is obtained, which is discretized in terms of space using a predictor-corrector scheme \cite{zago_2018}. The time step $\Delta t$ is chosen such that each particle $i$ fulfills CFL-like stability conditions shaped by the acceleration magnitude and speed of sound:
\begin{equation}\label{eq:sphcfltimestep}
    \Delta t_i\leq\text{min}\left\{0.3\sqrt{\frac{h}{|\mathbf{a}_i|}},0.3\frac{h}{c}\right\}.
\end{equation}
The presented framework is implemented in the modular open-source code GPUSPH \cite{herault_2010}. The C++ code is GPU-accelerated using CUDA and the computations are performed in single-precision. All simulations are carried out for 3D domains.

\section{Results}
The following section presents numerical results of the presented scheme for two example applications, and discusses their quality and the computational gain.

\subsection{Standing wave}
The literature frequently discusses the evolution of a standing wave in order to assess the capabilities of proposed numerical models \cite{antuono_2011,Colagrossi_2013,Macia_2011,zhang_2018}. Due to existing theoretical solutions for the kinetic energy of viscous fluids \cite{Lighthill_1978,Antuono_2013} the example of a standing wave offers a possibility to compare numerical results in a quantitative manner.

\smallskip In order to simulate a standing wave of wave height $H_s$, a basic water column is considered. The water column is of depth $d$ and the width is equal to the wave length $\lambda$ of the standing wave. Periodic boundary conditions are employed both in the $x$- and $y$-direction. Since the width is equal to the wave length, there will be exactly two antinodes and nodes.

The domain is discretized by a set of $n_p$ particles, which are placed on a regular grid with an interparticle spacing $\Delta p$ in the initial configuration at time $t_0=0$. The thickness of the numerical domain is chosen such that particles are not able to interact with themselves through the periodic boundary condition. Since the Wendland kernel with an influence radius of $2h$ is the employed kernel function, the thickness is set to $6h$, which is rounded up to the next multiple of $\Delta p$. If not noted otherwise, the interparticle spacing is set to $\Delta p=\tfrac{1}{4}\tfrac{H_s}{2}=1/256$, in order to fulfill the stronger criterion proposed by Antuono \cite{antuono_2011}, which demands a minimum resolution of four particles per wave amplitude.

The surface elevation is chosen as $\eta(x,t_0)=0$, such that the standing wave is in the state of maximum velocities at $t_0$. Accordingly, the initial values for the velocity yield
\begin{align}
    u_x\left(x,z,t_0\right)=\frac{H_s\text{g}k}{2\omega}\frac{\cosh\left(kz\right)}{\cosh\left(kd\right)}\sin\left(kx\right),\\
    u_z\left(x,z,t_0\right)=-\frac{H_s\text{g}k}{2\omega}\frac{\sinh\left(kz\right)}{\cosh\left(kd\right)}\cos\left(kx\right).
\end{align}
The pressure field is initialized according to the hydrostatic pressure term in Eq. \eqref{eq:pressureconti} as the dynamic pressure term vanishes for $\eta(x,t_0)=0$.

The dynamic boundary condition according to \cite{crespo_2007} is chosen as a boundary formulation for the static wall on the bottom, and the number of its particle layers is determined by the maximum radius of the neighbouring fluid particles.

For the sake of completeness it shall be noted that the prescribed initial conditions are constant throughout the $y$-axis.
\begin{table}[h!]
\caption[table caption text]{Standing wave parameters.}\label{table:standingwaveparam}
\centering
\begin{tabular}{l *{14}{c}}
$\lambda$ & $d$ & $T$ &$t_\text{end}$ & $H_s$ & g & $\rho_0$ & $\nu$ & $\gamma$ & $c_0$ & $\xi$\\\hline
1.0 & 1.0 & 0.8 & 20.0 & $1/32$ & 9.81 & 1000.0 & $10^{-6}$ & 7.0 & 89.0 & 0.1
\end{tabular}
\end{table}
Table \ref{table:standingwaveparam} displays physical and numerical values that were chosen for all standing wave simulations unless noted otherwise. The ratio $\tfrac{d}{\lambda}=1$ indicates deep water. The speed of sound $c_0$ is computed using the hydrostatic velocity $u_h=\sqrt{2\text{g}d}$ and yields $c_0\approx20\sqrt{2\text{g}d}$,
which limits the compressibility of the fluid. The period $T$ of the standing wave is obtained by considering $T=\frac{2\pi}{\omega}$, in which the angular frequency is obtained from Eq. \eqref{eq:dispersionrelation} and yields $\omega\approx7.854$ for given parameters. The artificial viscosity coefficient is dependent on the smoothing length $h$ and is determined according to Eq. \eqref{eq:nuofalpha}. For a standing wave, the Reynolds number can be estimated following
\begin{equation}
    \text{Re}=\frac{u_hd}{\nu}=\frac{\sqrt{2\text{g}d}d}{\nu},
\end{equation}
and yields Re$\approx4.4\cdot10^6$ for given parameters.
In order to compare obtained numerical results, the present work uses the solution \cite{antuono_2011}
\begin{equation}\label{eq:lighthill}
    E_\text{kin,th}=\epsilon^2\rho\text{g}\frac{\lambda d^2}{32}\text{e}^{-4\nu k^2t}\left(1+\cos\left(2\omega t\right)\right),
\end{equation}
in which the ratio $\epsilon=\frac{H_s}{d}$ is an estimate of the degree of linearity of the standing wave and yields $\epsilon=0.03125<<1$ in given case. The numerical counterpart of the kinetic energy is determined as a sum over all fluid particles in the domain:
\begin{equation}\label{eq:sphkineticenergy}
    E_\text{kin}=\sum_{j=1}^{n_p}m_j\frac{|\mathbf{u}_j|^2}{2}.
\end{equation}
In the following, the theoretical damping coefficient
\begin{equation}\label{eq:theoreticaldampingcoeff}
    \beta_0=-4\nu k^2
\end{equation}
will be of special interest. For given parameters, the theoretical damping coefficient yields $\beta_0=-1.579\cdot10^{-4}$. Within this section, the numerical counterpart of the damping coefficient will be computed as a least-squares exponential fitting of the numerical kinetic energy analogously to \cite{zago_2021}.

\subsubsection{Regularized CCSPH}
\begin{figure*}[h!]\label{fig:standingwaveetakinetic}
\centering
\subfloat[][Surface elevation over periods.]{
\includegraphics[]{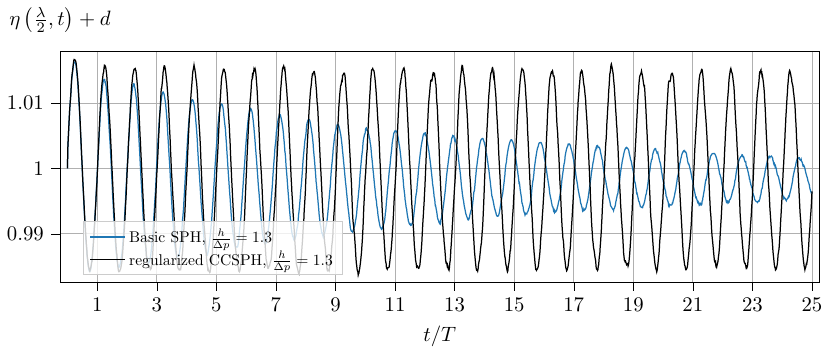}\label{subfig:v4surfaceovertime}}\\
\subfloat[][Kinetic energy over time.]{
\includegraphics[]{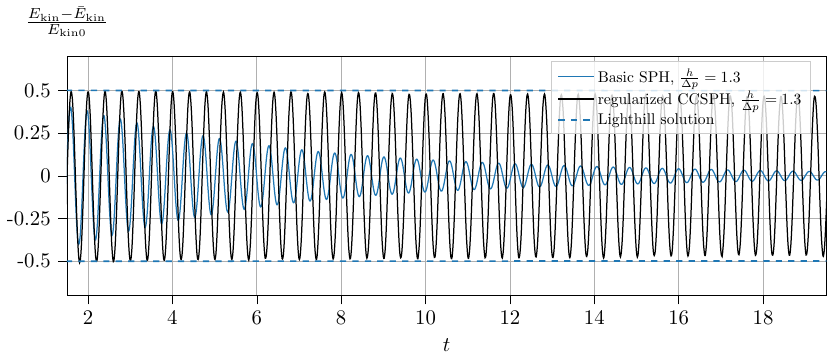}
\label{subfig:v4kineticenergyovertime}}
\caption{Results for the standing wave computed by three different SPH formulations and $h/\Delta p=1.3$.}
\label{fig:v4standingwave}
\end{figure*}

First, the general capability of the symmetric kernel gradient correction combined with the introduced weighting factors shall be demonstrated. For this purpose, it is applied to the entire domain except for the proximity of the boundary particles. Fig. \ref{subfig:v4surfaceovertime} displays the values of the surface elevation at a wave gage placed at $x=0.5$ and the kinetic energy of the standing wave over time for both a basic SPH framework, i.e. $B_{ij}=\mathbf{I}$, and CCSPH with the proposed regularization. For both formulations, a standard smoothing factor of $\tfrac{h}{\Delta p} = 1.3$ have been employed. Note that the evolution of the kinetic energy has been renormalized and by subtracting a one-period wide moving average, freed of its time-varying offset. The offset is due to a mixing and settling of particles after the simulation is started. This can be associated with the observation noted by Colagrossi et al. \cite{Colagrossi_2013}, that low-viscosity simulations with small smoothing lengths tend to form spurious vorticity.
The regularized CCSPH conserves the wave period almost exactly throughout the 25 periods. Basic SPH shortens the period and results in a phase shift of approximately $-1.275\pi$. In contrast, CCSPH slightly elongates the period and exhibits a phase shift of $+0.006\pi$ after 25 periods. 

Figure \ref{subfig:v4surfaceovertime} displays a decrease of the still water level for both formulations, which takes place early in the simulation and remains approximately constant afterwards. If the moving average of the surface elevation is considered, the basic formulation exhibits a decrease of approximately $-0.0015d$. The decrease of the still water level is smaller for the simulation in which CCSPH has been employed, namely roughly $-0.0004d$.

The most evident difference in the results is the conservation of amplitude, both in the surface elevation and in the kinetic energy. In contrast to basic SPH, regularized CCSPH is able to prevent large amounts of numerical dissipation, and yields an exponential decay coefficient of $\beta=-0.0037$ for the kinetic energy. This satisfying result correlates with the wave height, which is almost completely conserved after over 20 periods with $\tfrac{\bar{H}_{s,\text{end}}}{H_s}=0.9792$ close to $t_\text{end}$. If the results are compared to those of Zago et al. \cite{zago_2021}, it can be concluded, that the proposed regularization is a valid, parameter-free alternative to the employment of a threshold on the determinant of the corrective matrix.

\subsubsection{Restricting the corrected subset}
In order to optimize the efficiency of the framework, a restricted, corrected subset $\mathcal{C}$ has been introduced in Section \ref{sec:correctedsubsets}. If recalled, the condition \eqref{eq:sphpcondlambda} allows $\chi$ to define the depth of the fluid to which kernel gradient correction is applied to, as a proportion of the wave length. Since the wave length of the simulated standing wave and the depth of the water column are chosen as $\lambda=1$ and $d=1$, $\chi$ is equivalent to the depth and proportion of the entire domain. If $\chi=0.5$ is chosen, for example, the pressure condition for particles of subset $\mathcal{C}$ yields $p_i\leq4905$ for given parameters. Figure \ref{fig:standingwavechihalf} displays the particle distribution and different subsets computed by regularized CCSPH and $\chi=0.5$. Note that a large interparticle spacing of $\Delta p=1/64$ is chosen for the sake of illustration.

\begin{figure}[h!]
\centering
\includegraphics[]{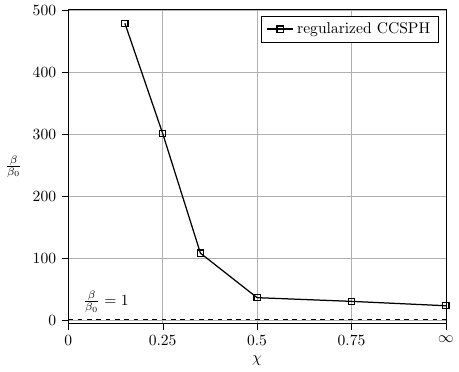}
\caption{Damping ratio over $\chi$ with fixed $\Delta p=1/256$.}\label{fig:dampingoverchi}
\end{figure}

In order to assess how the spatial restriction of kernel gradient correction impacts the results, Figure \ref{fig:dampingoverchi} displays the damping ratio over $\chi$. Note that $\chi=\infty$ suggests WSKGC being applied in the entire fluid domain except in the proximity of boundary particles, such as in the previous results.

If the result computed by the basic SPH model with $\tfrac{h}{\Delta p}=1.3$ in Figure \ref{subfig:v4kineticenergyovertime} is revisited, the absence of kernel gradient correction results in a damping ratio of $\tfrac{\beta}{\beta_0}\approx 987$. This result is obtained as the damping coefficient of a least-squares fit of an exponential function to the peaks of the kinetic energy. If regularized CCSPH is applied with $\chi=0.15$ and thereby only in approximately the upper 15\% of the domain, the damping ratio halves to $\tfrac{\beta}{\beta_0}\approx 479$.

The improvements maintain their significance until $\chi=1/2$, at which the damping ratio remains roughly constant until $\chi=\infty$. This is in good agreement with the theoretical contributions of water particles according to linear wave theory. Hence, kernel gradient correction is crucial if the particles are close to the surface, but negligible in depths of more than $\lambda/2$.

The simulations in which corrective coefficients were computed only for a restricted subset did not exhibit any irregularities at the border of the subsets or introduce other numerical problems. For the sake of completeness it shall be noted that the averaging of the corrective matrices applies to interactions between particles of different subsets as well and perhaps leads to a smoother transition between the two subsets.

Determining a sensible value of $\chi$ and thereby the location of this transition from one subset to another highly depends on the purpose of the simulation: If high accuracy is required, a greater value, such as $\chi=\tfrac{1}{2}$ is reasonable. The choice of this value can be adjusted according to the amount of wave lengths traveled by the simulated wave, or the amount of periods of interest. If many wave lengths or periods are simulated, $\chi$ should have a greater value to counteract the numerical dissipation, if not, it may be smaller in order to potentially spare computational effort.

The amount of lessened computational burden is characterized by the nature of the simulated wave. In case it is in shallow water, all particles in the domain will be members of the subset $\mathcal{C}$ for which corrective coefficients will be computed, and thus there will be no difference in the computational effort. If, however, the wave is either completely or partially in a deep water domain, the computational effort can decrease significantly. In the case of the investigated standing wave, the choice of $\chi=\tfrac{1}{2}$ led to an acceleration of roughly 25\% compared to CCSPH applied to the entire domain.

\begin{figure}[h!]
\centering
\includegraphics[]{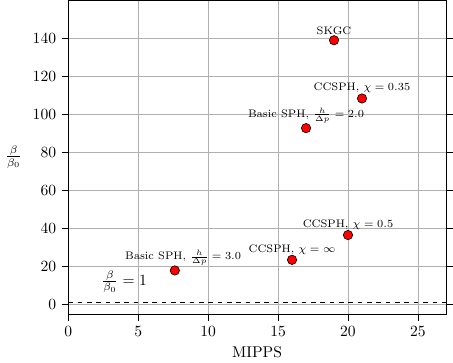}
\caption{Damping ratio over millions of normalized iterations per particle per second for a fixed $\Delta p=1/256$.}\label{fig:dampingoveripps}
\end{figure}

In order to assess how different values of $\chi$ impact the quality and the required amount of computational effort, Figure \ref{fig:dampingoveripps} displays the damping ratio over millions of normalized iterations per particle per second (MIPPS) for parameter sets that are sensible in practise. Note that the need for normalization stems from different time steps for different smoothing lengths according to Eq. \eqref{eq:sphcfltimestep}. All values have been normalized to the $\Delta t$ for $\tfrac{h}{\Delta p}=1.3$. The simulations have been run on two NVIDIA Titan Xp GPUs. For a frame of reference, the basic SPH model with $\tfrac{h}{\Delta p}=1.3$ yields 36 MIPPS on this system.

Large smoothing factors also impede the performance beyond the need for an evaluation of more particles within a neighborhood. Within this work, the total number of particles had to be increased in order to prevent particles interacting with themselves through the periodic boundary condition. Having this in mind, the conclusion that can be drawn from Figure \ref{fig:dampingoveripps} is emphasized. These performance issues add to the resulting irregular particle distributions highlighted in Zago et al. \cite{zago_2021}.

In the desired, lower right corner of the diagram, CCSPH with the proposed regularization can be found. This confirms that the weighted symmetric kernel gradient correction is not only highly satisfying with regards to the quality of the results, but it also maintains a reasonable amount of required computational effort. Restricting the application of regularized CCSPH to the part of the wave that contributes most to its kinetic energy provides the potential for further efficiency.

\subsection{Numerical results -- progressive wave train}\label{sec:progressivewave}
Most applications in coastal engineering involve progressive waves. In order to qualitatively assess the capabilities of the proposed formulation, a progressive wave train in a long wave tank is simulated.

\begin{figure*}[h!]
\centering
\includegraphics[]{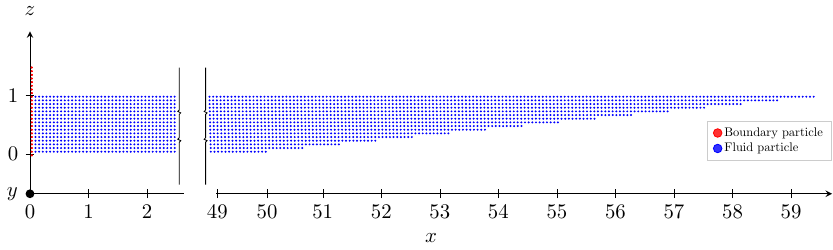}
\caption{Initial particle distribution of the numerical wave tank. For the sake of visualization, the interparticle spacing has been increased to $\Delta p=1/16$.}\label{fig:airywavesdistribution}
\end{figure*}

Figure \ref{fig:airywavesdistribution} displays the initial particle distribution of the numerical wave tank. The water depth is $d=1$, the thickness of the domain is set to $8h$, which is rounded up to the next multiple of $\Delta p$. The horizontal flat of the wave tank is $l=50$, which joins into a gentle slope with an inclination of 10\%, which reduces the impact of reflection.
As in the previously investigated example, periodic boundary conditions are employed along the $y$-axis. The boundary particles of the single flap wavemaker at $x=0$ act on neighbouring fluid particles according to the Lennard-Jones boundary formulation. They follow a prescribed motion which is derived from wavemaker theory for linear waves. The horizontal flat and the slope at the end are modeled using geometrical planes and the associated Lennard-Jones potential \cite{monaghan_1994}.

The initial velocities of the fluid particles are set to $\mathbf{u}_i=0$, and the initial pressure is calculated according to the hydrostatic pressure $p_i=\rho_0\text{g}z_i$.
\begin{table*}[h!]
\caption[table caption text]{Numerical wave tank parameters.}\label{table:numericalwavetankparam}
\centering
\begin{tabular}{l *{14}{c}}
$\lambda$ & $d$ & $T$ &$t_\text{end}$ & $H$ & g & $\rho_0$ & $\nu$ & $\Delta p$ & $\tfrac{h}{\Delta p}$ & $\gamma$ & $c_0$ & $\xi$\\\hline
1.5 & 1.0 & 0.98 & 68.11 & 0.1 & 9.81 & 1000.0 & $10^{-6}$ & 1/64 & 1.3 & 7.0 & 89.0 & 0.1
\end{tabular}
\end{table*}
Table \ref{table:numericalwavetankparam} displays physical and numerical parameters for the numerical wave tank. The nonlinearity parameter results in $\epsilon=\tfrac{H}{d}=0.1$. The ratio of the depth of the fluid and the wave length yields $\tfrac{d}{\lambda}=\tfrac{2}{3}$, and hence the wave can be classified as a deep water wave. As this is an example for a practical application, which involves a larger domain and accordingly greater number of particles, the smoothing factor will be limited to $\tfrac{h}{\Delta p}=1.3$ within this section. The ratio of wave height and interparticle spacing yields $\tfrac{H}{\Delta p}=6.4$, which fulfills the weaker criterion of Antuono et al. \cite{antuono_2011}. The simulation time $t_\text{end}$ ensures the required time $2T\frac{l}{\lambda}$ for the first wave front to propagate until the end of the horizontal flat has passed.

In the following, the results of CCSPH with the proposed regularization for the progressive wave train are qualitatively assessed. Analogously to the previous section, first the application of the formulation in the entire domain will be investigated, and subsequently the corrected subset will be restricted. Afterwards, the role of the weighting factors will be considered in the scope of this example.

The numerical wave tank poses a challenge to kernel gradient correction schemes. This is due to the bottom of the wave tank, which is modeled by geometrical planes without boundary particles. Accordingly, the surrounding fluid particles lack support. Especially the tip of the gentle slope at $x=59$ presents a particle distribution that is similar to neatly aligned particles and involves a singular corrective matrix. The weighting factors dampen this issue, but they do not solve it, as they are always greater than zero. As further treatment, it is a necessity to exclude fluid particles in the proximity of the boundary from the computation of corrective coefficients, in order to avoid spurious behavior. 

\begin{figure*}[h!]
\centering
\includegraphics[]{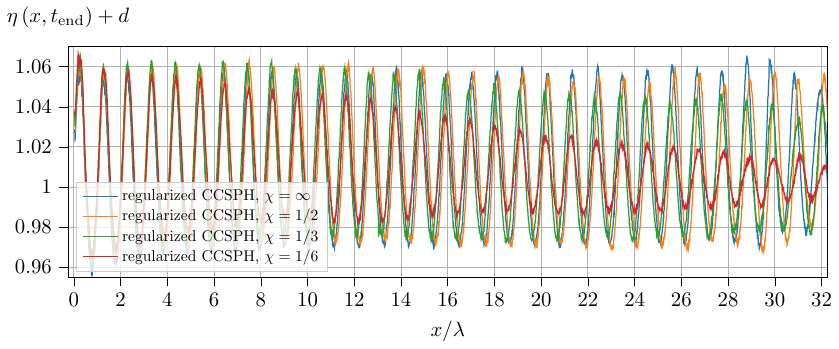}
\caption{Surface elevation of the progressive wave train at $t_\text{end}$, computed by regularized CCSPH with different values of $\chi$.}\label{fig:airywavesv4chi}
\end{figure*}

Figure \ref{fig:airywavesv4chi} displays the surface elevation in the numerical wave tank for results computed by WSKGC applied to the entire domain except for the proximity of boundaries for the time $t_\text{end}$. 
Regularized CCSPH applied in the entire water column except for the proximity of the boundary (denoted by $\chi=\infty$) is able to conserve the wave height well throughout the wave tank. The wave height remains roughly constant at approximately $0.9H$. It is possible that the difference of $0.1H$ is due to the efficacy of the employed wavemaker, as regularized CCSPH appears to maintain the wave height well. The deviations in the surface elevation for $x<3\lambda$ may be caused by the evanescent mode, which is created by the single flap wavemaker and acts in the proximity of the wavemaker. Further, the irregularities in the last three wave lengths might be due to minor reflections from the slope.

The behavior of the still water level is similar to the one displayed by basic SPH, but the still water level does not tend towards the initial water depth at the end of the wave tank. This may be due to the fact that the wave height is maintained throughout the domain, as the results for the basic SPH model suggest a correlation between wave height and still water level. The results of regularized CCSPH exhibit a maximum still water level at the end of the wave tank with an increase of roughly $1.2$\% compared to the initial water depth.

The wave length is notably shorter than the wave lengths predicted by the basic SPH model that ranged from $1.05\lambda$ to $1.125\lambda$. Nevertheless, it slightly elongates with values ranging from $1.01\lambda$ to $1.1\lambda$, resulting in approximately $1.6$ wave lengths less than the theoretical number of wave lengths throughout the wave tank.

In order to optimize the required computational effort, the corrective scheme is spatially restricted to particles close to the free surface, according to the criterion \eqref{eq:sphpcondlambda}. Figure \ref{fig:airywavesv4chi} also displays the surface elevation in the wave tank at $t_\text{end}$ computed by regularized CCSPH with different values of $\chi$.

Restricting the application of regularized CCSPH to fluid particles in proximity of the free surface of the progressive wave train yields results similar to the observed behavior for the standing wave. Great improvements are exhibited for values of $\chi$ up to $1/2$. Including further particles, that lay deeper than $\tfrac{1}{2}\lambda=0.75$, in the computation of corrective coefficients does not lead to significant differences in the results. Hence, the quality of results is rather similar for WSKGC being applied in the entire domain and it only being applied to particles that are closer than $\tfrac{1}{2}\lambda$ to the surface. This once more confirms the hypothesis made in Section \ref{sec:correctedsubsets}: Special treatment trough the proposed corrective method is crucial in regions of the wave, that contribute to its kinetic energy. Almost the entirety of those regions are located in depths up to half a wave length, which allows to refrain from the application of the corrective method in greater depths.

As mentioned above, the choice of $\chi=1/2$ only requires corrective coefficients to be computed in roughly 75\% of the fluid domain. This leads to an acceleration of almost 10\% compared to the application in the entire fluid body. However, it is evident, that depending on the required propagation distance, lower values of $\chi$ can be sensible choices as well. In this case, a choice of $\chi=1/3$ conserves wave energy almost as well as higher values of $\chi$ until a distance of $x/\lambda=16$. Especially in deeper water, the spared computational can be significant.

\section{Conclusion and outlook}
A localized application of a symmetric kernel gradient correction has been investigated. The proposed criterion for distinguishing particles that require the higher accuracy of kernel gradient correction from those that do not is motivated by linear water wave mechanics. The results reflect the theory very well: The impact of a higher-order method on the water wave energy is the greatest for particles close to the surface, and drops off exponentially over the water depth. 

This insight provides a useful engineering tool for costly simulations of water wave propagation. Depending on the required accuracy and propagation distance, the depth in which kernel gradient correction is applied to, can be adjusted. Especially for deep water waves, this can make long-term and large-scale simulations more feasible in terms of computational load. 

The present paper has been investigating linear water wave propagation. Nonetheless, the basic principle having a subset of particles with basic SPH approximations and a second subset of particles with a higher-order formulation, and assigning particles to either one of these subsets depending on certain properties, can be generic.


\end{document}